\documentclass{llncs}
\usepackage{graphicx}
\usepackage{fancybox}
\usepackage{url}
\def\systemname#1{\textsf{#1}}
\newcommand{\mizar}{\systemname{Mizar}}
\newcommand{\mml}{{\sc mml}}
\newcommand{\MML}{{\sc mml}}
\newcommand{\git}{\systemname{Git}}
\newcommand{\linux}{\systemname{Linux}}

\newcommand{\monotone}{\systemname{monotone}}
\newcommand{\bzr}{\systemname{bzr}}
\newcommand{\mercurial}{\systemname{mercurial}}
\newcommand{\arch}{\systemname{arch}}
\newcommand{\rcs}{\systemname{rcs}}
\newcommand{\cvs}{\systemname{cvs}}
\newcommand{\subversion}{\systemname{subversion}}

\newcommand{\ikiwiki}{\systemname{ikiwiki}}
\newcommand{\veedash}{\systemname{vdash}}
\newcommand{\sumusergroup}{{\sc sum}}
\newcommand{\envget}{\texttt{envget}}
\newcommand{\rsync}{\texttt{rsync}}
\newcommand{\dirtysandbox}{S_{d}}
\newcommand{\cleansandbox}{S_{c}}

\title{A Wiki for \mizar{}:\\Motivation, Considerations, and Initial Prototype\thanks{The final publication of this paper is
  available at www.springerlink.com}}

\author{Josef Urban\inst{1}\thanks{Supported by the NWO project
    \emph{MathWiki: A Web-based Collaborative Authoring Environment for
    Formal Proofs}.} 
  \and Jesse Alama\inst{2}\thanks{Partially
    supported by the European Science Foundation research project
    \emph{Dialogical Foundations of Semantics} within the ESF
    EUROCORES program \emph{LogICCC: Logic in the Humanities, Social
      and Computational Sciences} (funded by the Portuguese Science
    Foundation, FCT LogICCC/0001/2007).} 
  \and Piotr Rudnicki\inst{3}\thanks{Partially supported by NSERC.} 
  \and Herman Geuvers\inst{1}}

\institute{Radboud University, Nijmegen, the Netherlands \and New University of Lisbon, Portugal,
  Portugal \and University of Alberta, Edmonton, Canada}

\begin{document}

\maketitle

\begin{abstract}
  Formal mathematics has so far not taken full advantage of ideas from
  collaborative tools such as wikis and distributed version control
  systems (DVCS).  We argue that the field could profit from such
  tools, serving both newcomers and experts alike.  We describe a
  preliminary system for such collaborative development based on the
  \git{} DVCS.  We focus, initially, on the \mizar{} system and its
  library of formalized mathematics.
\end{abstract}

\section{Introduction and Motivation}

Formal mathematics is becoming increasingly well-known, used, and
experimented with~\cite{Hal08,harrison-notices}.  Verification of
major mathematical theorems such as the Kepler
Conjecture~\cite{Hal05}, the Four Color Theorem~\cite{Gon08}, and the
increasing use of verification for critical software and
hardware~\cite{KE+09,DKW08} are pushing the development of interactive
verification tools.  Indeed, there are already various online
repositories of formal proofs~\cite{afp-homepage,mml-homepage,ccorn}
and catalogs of deductive tools~\cite{qpq-homepage}.

The goal of the work presented here is to make formal mathematics 
\emph{accessible online} to interested parties by making the subject
widely available through online tools and interfaces.  We are
particularly interested in providing fast \emph{server-based} tools
for verification, web presentation, and collaborative development of
formal mathematics.

We believe that such tools are important for our field to attract
newcomers: they provide an attractive environment for exploring the
world of formal reasoning and the tools of the trade. The technology
we have in mind is vital also for existing members of the formal
mathematics community simply by making systems for formal mathematics
easier to access and use.  Ideally, the system should be so
straightforward that, to satisfy some momentary curiosity about a
formalism or a proof or a theorem, one could just visit a web page
instead of suffering through the installation of a new system.  In the
long run, we foresee a web-based repository of mathematics realizing
the vision of the QED Manifesto~\cite{qed-manifesto}.

Our effort has three principal features: it
\begin{itemize} 
\item is based on the notion of a wiki (understood here as a support
  for distributed, web-based collaborative project)
\item uses distributed version control system(s), and
\item uses server-based software to do the ``heavy
  lifting'' of verification and proof checking.
\end{itemize}
Let us briefly characterize these features of our approach.

The first main feature of our approach is the use of wikis.  Wikis are
well known; they offer collaborative web-based information resources.
The original wiki~\cite{c2-homepage} is nearly 15 years old.  This new
genre of web site has grown enormously.  Wikipedia is, evidently, the
most prominent example of a wiki: even casual Internet users are not
only aware of Wikipedia but use it often: links to Wikipedia pages are
often among the top results of web searches.

Mathematics is, thankfully, no exception to this trend.  Wikis and
other online repositories and communities for mathematics abound:
arXiv, MathOverflow~\cite{mathoverflow-homepage}, T.~Gowers'
PolyMath~\cite{polymath-homepage}, Wolfram
MathWorld~\cite{mathworld-homepage},
PlanetMath~\cite{planetmath-homepage},
ProofWiki~\cite{proofwiki-homepage}, etc.\footnote{Note that majority
  of these on-line services are non-profit efforts.}  These constitute
just a sample of mathematics' web presence; it does not seem feasible
to give a complete survey of the subject.  Yet, at present there are
no mathematical web sites for \emph{formal} mathematics that permit
collaborative editing and other online tools (though we know of one
prototypical effort~\cite{CorbineauK07}).  We aim to fill this gap.

The second main feature of our approach is the use of distributed
version control systems (DVCS).  Such systems are eminently suitable for
large collaborative efforts: they allow one to maintain repositories
of texts or code in the presence of multiple people working
independently or cooperatively.  DVCSs are becoming more widely used,
both in the commercial sector, in academia, and in the free software
community.  Our approach is novel because the application of DVCSs to
maintaining large bodies of formal mathematical proofs---objects that
are both computer code and human-readable text---is largely
underdeveloped.

The third and final main feature of our approach is the use of
server-based tools.  Like DVCSs and wikis, such systems are becoming
widely used.  The general reason for this is that the Internet is
becoming ubiquitous, faster, and more reliable.  Server-based
approaches can arguably help to attract newcomers to the subject
because it spares them the burden of locally installing formal
mathematics software and its associated libraries. Moreover,
computations that one might want to carry out are sometimes large and
should (or must) be carried out on servers rather than less powerful
client hardware.  In our case, proof checking and the generation of
rich semantic HTML presentations of formal proofs is, often, quite
expensive.

The rest of the paper is organized as follows:

\begin{itemize}
\item Section~\ref{use_cases} briefly describes the primary
  applications of a wiki for formal mathematics.
\item Section~\ref{dvcses} lists the essential features of DVCSs, and
  how we use them to provide robust and flexible back-end for our
  formal mathematical wiki.
\item Section~\ref{current_mizar_model} briefly discusses the current
  methods for developing the \mizar{} Mathematical Library (\mml) and
  identifies the main bottlenecks of the current model for massive
  distributed formalization efforts.

\item In Section~\ref{wiki_features} we discuss the features that a
  formal mathematical wiki should have and the requirements it should
  satisfy; we focus on the problem of maintaining the correctness of a
  formal library.

\item The initial implementation of the wiki for \mizar{}, based on
  the \git{} DVCS, is given in Section~\ref{prototype}. We explain the
  \git{} repository structure, the communication and division of work
  between the different repositories and branches, how we use \git{}
  hooks for guarding and updating the repositories, how we extract and
  recompute dependencies from the changed library and its
  re-verification and regeneration of its HTML presentation.

\item Possible future extensions are discussed in
  Section~\ref{future_work}.
\end{itemize}

\section{Use Cases}\label{use_cases}

We intend to provide tools for various collaborative tasks
in the formal mathematics %
communities. Our efforts should also aim to
include the whole mathematical community and be beneficial to it:
formal mathematics is a natural extension of traditional
mathematics in which the gaps, metaphors, and inaccuracies are
made ``computer-understandable'' and thus, in principle,
mechanically verifiable. The major use cases that we have in mind are
\begin{itemize}
\item public browsing of a human-readable web presentation of
  computer-verified mathematics;
\item facilitating the entrance to the formal mathematics community;
\item library refactoring (from small rearrangements of individual
  items and minor changes, to major overhauls);
\item authoring contributions, both large and small;
\item supporting the design of a formalization structure (concept
  formation, fleshing out definitions, bookkeeping postulates and
  conjectures);
\item offering tools to get help with simple proof steps;
\item gradually formalizing informal texts;
\item translating contributions between/among proof assistants,
  archives, and natural languages;
\item merging independently developed libraries.
\end{itemize}
In this paper, we narrow our focus to examine how some of the above
tasks apply to the \mizar{} system.  Thus, we are interested in the
aspects of the above problem that arise when working within a single,
fixed formal framework.  However, we keep in mind the rather grand
scale of the issues at hand.

\section{Towards distributed collaboration}\label{dvcses}

One of the most exciting but apparently unexplored topics in formal
mathematics is the use of distributed version control systems (DVCSs).
Such systems support tracking the development of work among authors
whose paths can be non-linear, proceeding down a variety of
routes/ideas that may or may not converge upon a final, conventionally
agreed-upon goal.

When thinking about DVCSs, a potential misunderstanding can arise that
we should correct.  One might get the impression that DVCSs simply
encourage chaos: without a central repository, who is to say what is
correct and what is not?  And would not DVCSs lead to a fragmented
community and wasted labor?

Although we can conceive such dystopian possibilities, we prefer
another point of view.  We want to emphasize the \emph{organized} in
\emph{organized chaos}, rather than the troubling noun.  It is helpful
to think of DVCSs not as a source of chaos, but rather as a tool for
putting some structure on the distributed efforts of a group of people
sharing a common interest.  In practice, DVCSs are used to organize
the efforts of many people working on crucial projects, such as the
\linux{} kernel\footnote{Just to give some feeling for the size, the
  compressed sources of the \linux{} kernel are roughly 53 MB; the
  compressed sources of the proofs in the \mizar{} Mathematical
  Library are 14 MB.}.  Although the DVCS model does not require a
central, standard repository to which everyone refers, there often are
strong conventions that prevent the disarray and confusion that can
come to mind when one first hears about DVCSs.  In the case of the
\linux{} kernel, for example, the entire community practically
revolves around the efforts of a single person---Linus Torvalds---and
a rather small group of highly trusted programmers.  A fairly wide
number of people still contribute to the \linux{} kernel, but in the
end, essentially all proposed contributions, if they are accepted at
all, pass through the core developers.  At least for the foreseeable future 
we propose to follow the same approach: a handful of experienced
\mizar{}ians\footnote{We thank Yuji Sakai for stressing the charm of this name.} will decide what constitutes the current official release
of the distributively developed system.

Foremost, we need to ensure that the current practices of developing
the \mizar{} Mathematical Library (\mml{}), which evolved over a
number of years, can still be supported.  Indeed, the core \mizar{}
developers can easily continue their current practices using DVCSs.
We would like to switch to distributed development in an incremental
fashion and we foresee deep involvement of the current \mizar{}
developers while switching to the new technology.

For our first pass at a formal mathematics collaborative tool, we have
chosen the \git{} system~\cite{git-homepage}.  \git{} is originally
developed by Linus Torvalds, the original author and primary
maintainer of the \linux{} kernel, for working with the large number
of contributions made to the \linux{} kernel.  \git{} enjoys
widespread use in various software communities (commercial, academic,
and open-source).

Our choice of \git{} is, to some extent, arbitrary.  A number of
widely used DVCSs are available (\monotone{}, \bzr{}, \arch{},
\mercurial{}) that could support the collaborative tasks we envision
for a formal mathematics wiki.  (Indeed, one project with aims similar
to ours--\veedash{}~\cite{vdash-homepage}---has proposed to use
\monotone{}.) %
We cannot, therefore, robustly defend our choice of \git{} except to
say that \emph{some} choice of DVCS must be made, and any other choice
of DVCS would likely be just as arbitrary.  We choose \git{} because
of our familiarity with it, and its wide usage in many projects that
we know of.

A full description of \git{} can be found on its
homepage~\cite{git-homepage} and the system can be learned through a
number of tutorials prepared by the \git{} community.  We believe that
\git{}'s concepts and operations are a good match for the kinds of
collaborative tasks that a formal mathematics wiki should support.
Here we limit ourselves to a skeletal presentation of \git{}'s main
features.

Like other version control systems, the \git{} system is based on the
notion of a \emph{repository}, a structured collection of objects
(texts, computer code, documentation, etc.) being worked on by members
of a team.  As a developer makes changes to the repository, a sequence
of \emph{commits} are made.  The sequence can split at various stages,
as when the developer or the community wish to take on a subproject or
otherwise pursue some line of investigation.  The full history of a
repository can thus be more aptly understood as a tree rather than a
simple sequence of changes.

A single developer can either start a new project afresh, or can
\emph{clone} the repository of some other developer working on the
project of interest.  Unlike traditional non-distributed version
control systems such as \rcs{}, \cvs{} and \subversion{}, \git{}
repositories (and those of some other DVCSs, too) are ``complete'' in
the sense that the developer has essentially unlimited access, even
when offline, to the entire history of a repository.  When online, the
developer can share his changes with others by \emph{pushing} up his
changes to another repository (or repositories), and he can stay
connected to his colleagues by \emph{pulling} down their changes.

The \git{} system also provides \emph{hooks} that can be used to
implement various guards and notifications during the various
repository actions (\emph{commit}, \emph{push}, etc.). This can be
used to allow only certain kind of changes in the repository. For
example, our initial wiki prototype for Mizar (See
Section~\ref{prototype}) makes use of suitable hooks to ensure that
the updated articles and the whole updated library repository are
always in a correct state.

In our initial prototype, we introduce a publicly accessible, central
repository---in the spirit of Wikipedia or the main \linux{} kernel
branch---that serves as a correct and verified snapshot of the current
\MML.  Our tools are targeted at public editing of the repository,
while ensuring the library's coherence\footnote{Coherence of \mml{}
is closely related to formal correctness and is a narrower notion
than integrity of \mml, cf.~\cite{RudnickiT03}.} and correctness as the changes
are made.

\section{A special case: the \mizar{} developers}\label{current_mizar_model}

Among the first targets for our implementation are the core \mizar{}
developers.  It will be worthwhile, then, to explain how \mizar{} is
currently developed.

The history of the \mizar{} project is briefly presented
in~\cite{MatuszewskiR04a} and a brief overview of the current state of
the project can be found in~\cite{NaumowiczK09}. 

The development of the \mizar{} Mathematical Library (\mml{}), the
principal, authoritative collection of \mizar{} contributions (called
``articles''), has been the main activity of the \mizar{}
project\footnote{\url{http://mizar.org}} since the late 1980s, as it
has been believed within the project team that only substantial
experience may help in improving (actually building) the system.  An
article is presented as a text-file and contains theorems and
definitions.  (At the time of writing, there are 1073 articles in
\mml{}, containing 49548 facts/theorems and 9487 definitions.)  The
articles of \mml{}---in source form---are approximately 81MB of data.

The current development of \mml{} is steered by the Association of
\mizar{} Users (in Polish: Stowarzyszenie U\.{z}ytkownik\'{o}w Mizara,
abbreviated \sumusergroup{}), which owns the copyright to the library.
\sumusergroup{} appoints the \mml{} Library Committee---the body
responsible for accepting articles, arranging reviews, and maintaining
official releases of the library.  \mml{} is organized in an
old-fashioned way, resembling printed repositories of mathematical
papers. Since \mml{} articles exist electronically, it is relatively
easy to re-factor the \mml{} contents by, say, changing items such as
definitions and theorems, or by deleting, or by moving them.  After
such modifications, the \mizar{} processor is run on the entire \mml{}
to verify its coherence.  Currently, the Library Committee does the
greatest amount of refactoring of \mml{}, which includes the following
activities:
\begin{itemize}
\item updating existing articles to take advantage of changes in the
  \mizar{} language, the \mizar{} processor, and changes in \mml{};
\item moving library items where they more appropriate locations;
\item building an encyclopedia of mathematics in \mizar{} which
  consists of creating new articles by moving topically related items from all
  over the \mml{} into one place;
\item generalizing definitions and theorems;
\item removing redundant items; and
\item merging different formalizations of the same material.
\end{itemize}
This process is under control of the Head of the Library Committee;
about three people do most of the revisions. It can also happen that
the original \mizar{} authors re-factor their own past contributions
after they notice a better way to formalize the material.  The
remaining \mml{} revisions are usually suggested by posting to the
\mizar{} mailing lists, or mailing directly to someone on the Library
Committee.

Information about what each of the \mml{} refactorers is doing is not
broadcast in any widely accessible way. The typical method is through
an email announcement: ``I am now formalizing XYZ,
so please wait for my changes''. Such announcements are frequently
accompanied by a solicitation for remarks and discussion.

Such a process for collaboratively editing \mml{} can be
problematic and is far from ideal.  The main problem with the present
method is that it does not scale up: although it can be used with some
success among few people, we imagine a much larger community where the
collaboration problem surely cannot be effectively solved by the
present approach.
Moreover, new users face an unpleasant experience of developing a new
article (or a group of interrelated articles) and, before finishing
their work, they learn of a newer, substantially different official
\mml{} version that is (partly) incompatible with their working
articles.  Updating such an unfinished development can be quite
time-consuming (but the Library Committee offers help in such cases).

\section{Formal wiki features and issues}\label{wiki_features}

The experience of \mizar{} authors is that while developing some new
formalizations they notice that many \mml{} items can be improved by,
say, weakening the premise of an existing theorem or widening the type
of a parameter in a definition.  Ideally, an author should be able to
introduce such small changes to the repository instead of asking the
Library Committee and waiting for a new version of \mml to be
released.  With a DVCS such small changes can be incorporated into a
daily build as there are mechanical means for verifying the
correctness of such edits.

For the benefit of users, there must be something like an official
\mml{} version.  We believe that with the DVCS development model, such
official versions can be produced less frequently, subsequent official
versions could differ more substantially, and all differences could be
documented.

We foresee the following goals when treating \mml{} as a wiki:
\begin{itemize}
\item content \emph{correctness};
\item \emph{incremental} editing; and
\item \emph{unified presentation} of the library for browsing.
\end{itemize}
Finding the right trade-off among these goals is not trivial. Allowing
incremental editing of a formal text can lead to its incorrectness. On
the other hand incrementality is an important feature of wikis. Thus
the formal wiki should be able to save one's work even if it is a
major rewrite of the existing code, and not completely correct. But
the library is also a source of information for other users/readers
and its current stable/correct version should be well presented. This
means, however, that all kinds of unfinished work (active
works-in-progress, abandoned attempts) should be reasonably hidden
from the readers who just came looking for information.

Thus it seems that we have to combine the methods used for distributed
software development---allowing many people to work on different
parts, possibly breaking things, and easily accessing and merging each
other's work--with the methods and mechanisms used for development of
informal wikis like Wikipedia, where the content is primarily intended
for human reading, and things cannot be ``formally broken'' too much
(obviously, large projects like Wikipedia do have some automated
structural checks, but evidently most of the review is done by
humans).  As of now we are only collecting experience on how to build and maintain 
a wiki of contents with formally defined semantics and correctness.  It is not clear 
to what degree issues raised in designs of less formal repositories are
relevant to our task.

\subsection{Degrees of Formal Correctness and Coherence}

The important difference between text-based wikis like Wikipedia and
formal libraries like \mml{} is the strong, well-defined notion of
\emph{formal correctness}. As regards \mml{}, one can consider
several aspects/scopes of formal correctness that are not available
(not even in principle) for traditional wikis such as Wikipedia.

To begin, consider the notion of \emph{link coherence} on, say,
Wikipedia.  Suppose someone edits only the body text of the Wikipedia
entry for the Eiffel Tower.  The article's internal coherence is
unchanged, and the link coherence of Wikipedia as a whole is
unaffected by the edit.  However, other kinds of edits are permitted:
\begin{itemize}
\item if one deletes the Eiffel Tower article or renames it, then
  (part of) Wikipedia becomes link incoherent because some internal
  links become broken.
\item if one deletes or renames a section of the Eiffel Tower article,
  then, as in the first case, it is quite possible that other
  Wikipedia articles become ``tainted'' by this edit by containing
  links to a subsection that no longer exists.
\end{itemize}
Internal link coherence can be enforced in various ways.  Some methods
are entirely computational (that is, can be carried out autonomously
by a machine and require no human intervention); others depend in part
on the (human) Wikipedia community.  Further, this notion of coherence
can be enforced by responding to potentially problematic edits, or by
simply not giving users the means to make them.  For example, the
problem of deleting pages can be addressed by simply disallowing
deletions.  If such edits are allowed (perhaps only by certain users),
an autonomous wikibot, constantly patrolling Wikipedia, can enforce
internal link consistency by updating every Wikipedia article that
refers to the Eiffel Tower article.  This kind of bot-based approach
could also respond to internal section renaming or deletion.  Finally,
internal link coherence can be enforced by Wikipedia's editor's tools,
which give humans an environment for editing that does not
straightforwardly permit such section renaming in the first place.

A wiki based on formal texts like \mizar{} articles permits one to
define the notion of \emph{change propagation}.  When a user changes a
piece of the \mizar{} library, we must take into account how this change 
affects other parts of \mml. For example, if
out of curiosity one changes the definition of the constant $\tau = (1
+ \sqrt 5)/2$ to, say, $1/2$, then, obviously, this edit will have
some effect on the library (some theorems about Fibonacci numbers that
involve this classical ratio will be invalidated).  Likewise, if one
rearranges the theorems or definitions in an article, this too can
affect parts of the library.  On the other hand, some edits are safe:
one can append new content to the end of existing articles without
affecting the library's coherence.

The problem is that some modifications preserve the coherence of the
wiki, but some do not.  The typical task we face in maintaining the
coherence of \mml{}, considered as a wiki, is that a user has
changed or added some articles to the library, and we want to verify
that these changes are admissible, that is, that they maintain the
coherence of \mml.  In extreme cases, checking the admissibility
of a user's edit can, conceivably, require re-verifying of the entire
library (for example, imagine changing the the fundamental
set-theoretical axioms on which \mml{} rests).  The cost of such
admissibility checks is correspondingly large.  However, there are a
number of more or less typical cases that can be dealt with less
expensively.  In the next section we explain how we have implemented
these checks.

\section{Prototype}\label{prototype}

\subsubsection{Suitability of DVCSs and wikis\\}  

A natural object of interest when thinking about wikis and DVCSs
together are wikis based on DVCS.  According to a recent
listing,\footnote{\url{http://git.wiki.kernel.org/index.php/InterfacesFrontendsAndTools}}
there are today more than a dozen wikis based on the \git{} DVCS or
using \git{} as a back-end.  Our design decision about wiki for
\mizar{} is: the data (articles) in the \git{} repository should
always be self-sufficient (independent of the wiki functionalities),
and easily exchangable between various installations (that can be used
just for local development, as a back-end for another wiki-like
system, etc.).  Wikis form a dynamic field of interest, so tight
coupling the \mizar{} data to a specialized and possibly not widely
adopted wiki storage implementation could cause difficulties for
developing alternatives and for integrating formal mathematics into
other, larger frameworks.  All such alternatives should provide
convenient communication at the DVCS (data) level with others. This
seems to be a reasonable invariant for building all kinds of web sites
and tools for creation of formal mathematics (and probably for
collaborative creation of code in general).

\ikiwiki{} is one of the most developed an probably best-known
\emph{wiki compilers}.  Wiki compilers process a file or set of files
written in a special (usually simplified) syntax and turn them into
static HTML pages. It is possible to build an initial prototype of a
wiki for \mizar{} by customizing \ikiwiki. We have explored this path
and are aware of many functionalities of \ikiwiki{} useful for our
task. We have also considered the peculiarities that make our domain
different from the typical (one-page-at-a-time) paradigm of such wiki
compilers, and decided to gradually re-use and plug-in the interesting
\ikiwiki{} functionalities, rather than trying to directly customize
the large \ikiwiki{} codebase to our nonstandard task.

\subsubsection{Prototype Overview:}
\begin{figure}[tpb]
    \includegraphics[width=13cm]{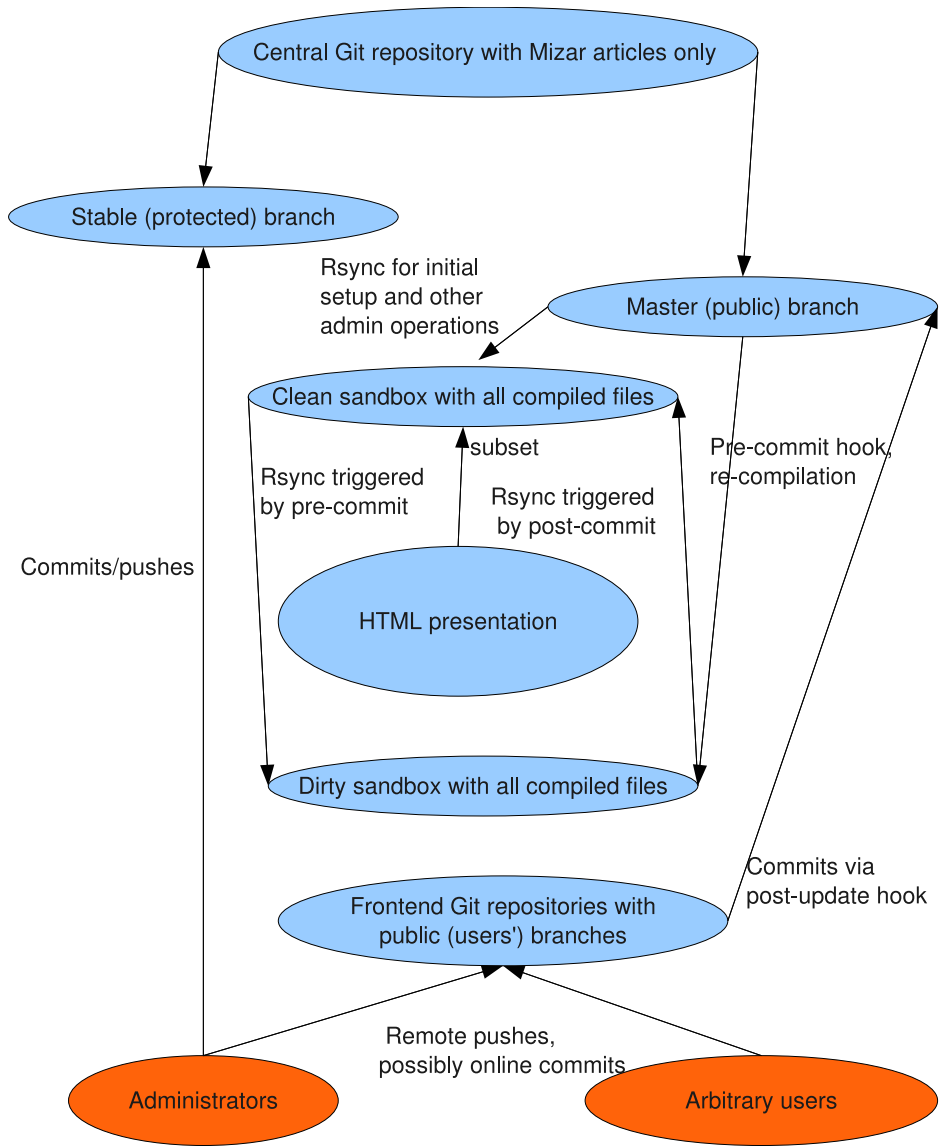}
  \caption{Mizar wiki structure} 
  \label{Repos}
\end{figure}
As we discussed earlier, our prototype (see
Figure~\ref{Repos}) %
is initially focused on the \mizar{} developers who edit their work
mainly offline and submit their changes to be viewed
online\footnote{\url{http://mws.cs.ru.nl/mwiki}}
by other developers. The repository structure (and suggested future
features) of our prototype is as follows:\footnote{The code is
  available from \url{http://github.com/JUrban/mwiki}}
\begin{itemize}
\item There is a central repository on our server with the
  \emph{stable} branch, and the \emph{master} branch. This repository
  can be (using post-commit hook) kept in sync with a version
  published at, say, GitHub (\url{http://github.com}, a high-profile
  web site for collaborative \git{}-based development) or other easily
  accessible public \git{} repository web sites. Anybody can clone the
  published copy from such public locations, saving us some of the
  responsibility for maintaining access and sufficient bandwidth to
  our repositories.
\item The repository contains basically only the source \texttt{.miz}
  files (\mizar{} formalizations), just as the \linux{} repositories
  only contain the source \texttt{.c}/\texttt{.h} files. All files
  that depend on the sources, and possibly intermediate meta-data
  constructible from them, are not part of the repository, and are
  excluded from the repository.  We achieve this using \git{}'s
  \texttt{.gitignore} feature.
\item The \emph{stable} branch of the repository is only available to
  administrators. New official versions of \MML{} are pushed there.
\item The \emph{master} branch is the default one and it can be
  updated by any formally correct change (guarded by a suitable \git{}
  pre-commit hook).  Essentially anybody can make changes (this is the
  wiki feature) using the mechanisms and checks described
  later. The \emph{stable} branch and the \git{} history should serve
  as tools for easily reverting bad changes.
\item The central repository is not directly updatable by
  users. Instead, it is cloned to another (frontend) repository on the
  server to which \git{} remote pushes can be made without any
  correctness checking. Limited control should be exercised to
  prevent malicious use\footnote{It is possible to have a
    number of such frontends, and with sufficient infrastructure in
    place to actually move the main non-verifying frontends again to
    public hubs like GitHub.}. Upon a successful remote push to the
  frontend, the \git{} post-receive hook is triggered. This hook
  attempts a commit to the master branch of the central repository,
  triggering in turn its pre-commit hook, and the formal verification
  of the updated library.
\item Upon successful commit into the central repository, a
  post-commit hook is triggered. This hook generates HTML for the
  updated library, publishes it on the web, and does possible further
  related updates (updates of related clones on GitHub, notifications,
  etc.)
\item The {\tt gitweb} graphical web interface can be used for
  browsing the repository (comparing different changes and versions,
  and watching its history, etc.). Alternatives to {\tt gitweb}
  abound: one is to use all the \git{}-related tools maintained and
  developed for GitHub.
\item Editing is, initially, done locally, using a local customized
  editor like Emacs (used by most current \mizar{} users today);
  later, the articles are submitted using \git{} mechanisms. Similar
  ``offline editing'' is not uncommon in the wiki world: in the Emacs
  environment, for example, there are tools for similar offline
  interaction with Wikipedia. There are a number of options for
  providing additional direct in-browser editing (useful for smaller
  edits), ranging from the basic text-box and submit button to more
  advanced general web-based editors like in Wikipedia, to specialized
  online mathematical editors like ProofWeb.
\end{itemize}

\subsection{Prototype Implementation}

The implementation of the pre-commit checks, the post-commit cleanup,
and related infrastructure is based on the following.  We make
essential use of makefiles to specify which files need to be rebuilt,
how to build them, and how they depend on other files.  In the central
repository we have one central makefile $M$ and, for each article $a$,
a makefile $M_{a}$ that specifies on which other articles $a$ depends
(e.g., what notations, definitions, theorems $a$ uses).  The master
makefile $M$ has targets that ensure that the whole of the
(about-to-be-submitted) \MML{} is coherent and without errors.
Naturally, when one submits a small change, it is generally not
necessary to re-verify the entire \MML{}, and the makefiles are
designed to re-verify the minimal necessary set of
dependencies. The verification is carried out in a ``fully compiled''
\MML{}, so all the auxiliary results of previous verifications
(analogous to \texttt{.o} files for GCC) are available, from which the
make program can infer whether they need to be re-computed.

As the library changes, the dependencies among the articles can
change, and re-writing the makefiles by hand each time a dependency
changes would be tedious. Following tools like {\tt makedepend} for C
and other languages (and probably some similar tools for other proof
assistants), we have created the \envget{} tool based on the \mizar{}
codebase for quickly gathering the dependencies that are explicitly
declared by an article\footnote{We are making use of the fact that in
  each verifiable \mizar{} article, one must declare what kinds of
  notations, definitions, proofs, etc., one depends on.  If \mizar{}
  did not have this feature, then calculating dependencies would,
  presumably, be more difficult. Also note that these are just
  dependencies between articles, while it is certainly interesting
  future work to also calculate the precise dependencies between
  smaller-scale article \emph{items} and use this information for
  smarter and leaner re-verification. The MPTP
  system~\cite{Urban06-jar} can be used for extracting information of
  this kind.}. Thus the makefiles $M_{a}$ for each article $a$ are
themselves generated automatically using makefile targets defined in
the master makefile $M$.  The XML output of \envget{} is combined with
a suitable XSL stylesheet, producing $M_{a}$ for an article $a$,
containing only one target, and specifying the articles on which $a$
depends. These dependency makefiles are refreshed only when
the corresponding article is changed. This leads to a reasonably
efficient Makefile-based setup:  only the dependencies of changed files
get possibly changed, and if that happens, only the (dynamically
changed) dependency subgraph of \MML{} influenced by the committed
changes gets re-verified.

Where is the make program (governing verification of changes) invoked?
How do we ensure that the central repository, assumed to be a coherent
copy of the \mizar{} distribution, does not get ``tainted'' by
incoherent user updates?  In addition to a fully-compiled, coherent
clean \mizar{} ``sandbox'' directory $\cleansandbox$, we maintain a
(possibly) dirty sandbox directory $\dirtysandbox$ that may or may not
be aligned with $\cleansandbox$.  The two directories vary in the
following manner.  When a new user commit is proposed, we use \rsync{}
tool~\cite{tridgell-thesis} to efficiently synchronize
$\cleansandbox$ and $\dirtysandbox$, thereby ensuring that it is
clean; we then copy all new \mizar{} source files to the dirty sandbox
(that was just cleaned). Note that using \rsync{} for this task
provides a reasonable trade-off for solving several different
concerns:
\begin{itemize}
\item to check the newly arrived changes without possibly destroying
  the previous correct version, we need to have a fresh directory with
  the most recent correct version;
\item the directory should contain as much pre-computed information as
  possible, because re-verifying and HTML-izing the whole library
  (more than 1000 articles) from scratch is an expensive operation
  (taking hours, even with high-end equipment), while we want to be as
  real-time as possible and re-use all available precomputed information;
\item while the directory containing just the \mizar{} articles is
  reasonably small (81MB at the moment), the directory with the
  complete pre-computed information is prohibitively big: the internal
  library created from the \mizar{} articles, the environment for each
  article, and the HTML files are together more than
  10GB.\footnote{This blow-up is caused by creating a local (XML)
    environment files for every article, and by having detailed XML
    and HTML representations of the article. A lot of information is
    kept in such files in a verbose form.}  Simply copying all this
  pre-compiled information would rule out a real-time experience,
  especially in cases when the user wants to change an article on
  which not many other articles depend.
\end{itemize}
Using \rsync{} into a fresh directory addresses the first two
issues (keeping the clean install intact and not re-verifying all
from scratch).  Using \rsync{}, on average, reasonably ensures (by relying
on smart \rsync{} mechanisms) that of the
clean sandbox over the dirty one does not take too long.

Since the clean sandbox $\cleansandbox$ contained
\emph{everything}---\mizar{} source files and all generated auxiliary
files---the dirty sandbox $\dirtysandbox$ now contains (after adding
the newly modified \mizar{} files) nearly everything, too.  All that is
missing are up-to-date auxiliary files\footnote{That is, files
  witnessing the correctness of the verification, updated library
  files, re-generated HTML files, etc.} to be generated from the
modified \mizar{} source files.  For that we invoke the master
makefile $M$ to generate possibly new dependency makefiles $M_{a}$,
and then we invoke the master makefile to request a
re-build/re-verification of the entire \MML.  Since we are working in
a sandbox $\dirtysandbox$ that contains the all results of previous
builds of the entire \MML{}, ``most'' of the \mizar{} source files
will not need to be looked at.  (We put ``most'' in quotes because if
one proposes to commit an update to a sufficiently basic \mizar{}
article, then a good deal of computation is required to verify and
propagate the change over the whole library, even though it could
actually be a rather small edit.  But such ``foundational'' edits are,
apparently, uncommon.)

By using makefiles this way we are also able to exploit their ability to run
multiple jobs in parallel.  The dependency graph of the full \MML{} is
wide and rich enough that, when making ``normal'' edits, we can
indeed benefit from this feature of make when running on multi-core machines or 
when employing grid/cloud-computing.
Indeed, this is crucial for us to provide a sufficiently
quick response to users who submit ``normal'' changes to \MML{}.

Further opportunities for parallelization, based on a finer-grained
analysis of \mizar{} articles than the one discussed here (where the
dependency relation is between an entire \mizar{} article on other
articles), are being investigated.

\section{Future Work and Summary}\label{future_work}

A number of features can be added to our prototype. As already
discussed, the world of wikis is dynamic and it is likely that all
kinds of web frameworks and technologies will be used in the future
for presenting the \mizar{} library, and collaboratively growing
it. This is also why we stress a good basic model for dealing
with the main data (articles) and processes (collaboration), i.e.,
using \git{} as a good mainstream DVCS, with a rapidly growing number
of tools, frontends, and public collaborative platforms based on it.

The features that are being implemented at the time of writing this paper
include: basic web-based editing; finer
itemization, dependencies, parallelization and thus verification and
HTML-ization speed for both simple (one-article) commits and
multi-article commits; plugging in all kinds of powerful proof advice
tools (automated theorem provers, library search engines, AI systems,
etc.). Obviously these require nontrivial effort on all these tools
and interfaces to them. Nontrivial and continuing work is actually
already providing good HTML presentation of the formal articles, with
all kinds of additional useful information and functions (following
the symbols to their definitions being the most obvious one).

We have already mentioned the variety of tools available for \git{}
and how public collaborative sites like GitHub rapidly develop all
kinds of collaborative functions on top of \git{}. A similar remark is
also true about, e.g., \ikiwiki{}, and possibly about other wikis
based on DVCSs. Thus, it seems to us that the mission of formal
mathematical wiki builders should be to watch these rapidly developing
collaborative technologies, customizing them (by providing suitable
commit hooks, HTML-izers, dependency utilities, change propagation
models, etc.) and complementing them (by providing all kinds of
support tools developed for formal mathematics) rather than competing
with them by starting from scratch.

While this paper is mainly about the \mizar{} proof assistant and its
library, it should be clear that the functionalities we discussed
(tools for good dependency extraction and focused re-verification and
HTML-ization, possibly itemization and parallelization, all kinds of
useful proof advice tools, etc.), the ideas and work presented here
could be instantiated also to other proof assistants (Coq, Isabelle,
HOL, etc.). When this is done, formal wikis could become the place
where various formalisms and alternative formalizations meet, allowing
collaboration on large formal projects, possibly having mechanisms
also for gradual formalization of informal mathematical texts, and
also allowing formal texts that are not completely correct. Our hope
is that this infrastructure will attract more ``normal/traditional''
mathematicians to the advantages of formal mathematics, gradually
making it more digestible, and possibly thus allowing further
important steps in the inevitable computerization of human
mathematics.

\bibliographystyle{splncs03}
\bibliography{bib1}
\end{document}